# Electrocardiographic Deep Learning for Predicting Post-Procedural Mortality

David Ouyang[1,2], John Theurer[1], Nathan R. Stein[1], J. Weston Hughes[3], Pierre Elias[4,5], Bryan He[3], Neal Yuan[1], Grant Duffy[1], Roopinder K. Sandhu[1], Joseph Ebinger[1], Patrick Botting[1], Melvin Jujjavarapu[1], Brian Claggett[6], James E. Tooley[7], Tim Poterucha[4], Jonathan H. Chen[8], Michael Nurok[9], Marco Perez[7], Adler Perotte[4], James Y. Zou[3,10], Nancy R. Cook[11], Sumeet S. Chugh[1,2], Susan Cheng[1] and Christine M. Albert[1]

1. Department of Cardiology, Smidt Heart Institute, Cedars-Sinai Medical Center
2. Division of Artificial Intelligence in Medicine, Department of Medicine, Cedars-Sinai Medical Center
3. Department of Computer Science, Stanford University
4. Milstein Division of Cardiology, Department of Medicine, Columbia University Irving Medical Center
5. Department of Biomedical Informatics, Columbia University Irving Medical Center
6. Division of Cardiovascular Medicine, Department of Medicine, Brigham and Women's Hospital
7. Division of Cardiology, Department of Medicine, Stanford University
8. Division of Bioinformatics Research, Department of Medicine, Stanford University
9. Division of Anesthesia, Department of Surgery, Cedars-Sinai Medical Center
10. Department of Biomedical Data Science, Stanford University
11. Division of Preventive Medicine, Department of Medicine, Brigham and Women's Hospital

Correspondence: David.ouyang@cshs.org and Christine.albert@cshs.org

**KEY POINTS**

**What is new?**

- Deep learning approaches applied to pre-operative electrocardiograms (ECGs) can offer additional information that could improve discrimination of post-operative mortality and improve upon current pre-operative risk stratification tools.
- PreOpNet is a deep learning algorithm that leverages ECG waveform signals to augment discrimination of post-operative mortality, and its performance was evaluated in three independent healthcare systems.

**What are the clinical implications?**

- ECGs contain significant additional information that can assist in pre-operative risk-stratification and deep learning evaluation of ECGs can better discriminate post-operative mortality than the Revised Cardiac Risk Index.
- The deep learning algorithm worked equally well for risk stratification of cardiac surgeries, non-cardiac surgeries, and catheterization lab procedures.
- Further research is needed to evaluate the value of implementing deep learning assessment of pre-operative ECGs for risk stratification.

**ABSTRACT (word limit 250)**

**Background.** Pre-operative risk assessments used in clinical practice are limited in their ability to identify risk for post-operative mortality. We hypothesize that electrocardiograms contain hidden risk markers that can help prognosticate post-operative mortality.

**Methods.** In a derivation cohort of 45,969 pre-operative patients (age 59±19 years, 55% women), a deep learning algorithm was developed to leverage waveform signals from pre-operative ECGs to discriminate post-operative mortality. Model performance was assessed in a holdout internal test dataset and in two external hospital cohorts and compared with the Revised Cardiac Risk Index (RCRI) score.

**Results.** In the derivation cohort, there were 1,452 deaths. The algorithm discriminates mortality with an AUC of 0.83 (95% CI 0.79-0.87) surpassing the discrimination of the RCRI score with an AUC of 0.67 (CI 0.61-0.72) in the held out test cohort. Patients determined to be high risk by the deep learning model's risk prediction had an unadjusted odds ratio (OR) of 8.83 (5.57- 13.20) for post-operative mortality as compared to an unadjusted OR of 2.08 (CI 0.77-3.50) for post-operative mortality for RCRI >2. The deep learning algorithm performed similarly for patients undergoing cardiac surgery with an AUC of 0.85 (CI 0.77 – 0.92), non-cardiac surgery with an AUC of 0.83 (0.79 – 0.88), and catherization/endoscopy suite procedures with an AUC of 0.76 (0.72 – 0.81). The algorithm similarly discriminated risk for mortality in two separate external validation cohorts from independent healthcare systems with AUCs of 0.79 (0.75-0.83) and 0.75 (0.74-0.76) respectively.

**Conclusion.** The findings demonstrate how a novel deep learning algorithm, applied to pre-operative ECGs, can improve discrimination of post-operative mortality.

**INTRODUCTION**

In the United States, over 20 million surgeries are performed annually.[1] Pre-operative risk assessment for adverse procedural outcomes – the most serious of which include death – is routinely performed in clinical practice.[2,3] However, current approaches to predicting post-operative risk remain limited.[4-6] Over the past three decades, expert guidelines and tools for facilitating pre-operative assessments have evolved to include biomarkers in addition to demographic and clinical data.[2,7-10] However, even the most comprehensive risk scores based on recognized risk markers provide only modest ability to discriminate post-operative outcomes with areas under the curve (AUC) ranging from 0.57 to 0.75.[6,11-14]

Early identification of patients at high risk for post-procedural mortality can significantly guide patient care, consideration of alternative treatment pathways, and aid in shared decision and communication of risk to patients[15,16]. Post-operative biomarker levels have been used to further risk stratify for post-operative mortality with strong discrimination[17-19], however the dependence on post-procedural information limits the opportunity to risk stratify pre-operatively. Novel methods of perioperative risk assessment are needed to achieve better discrimination across the heterogeneous population of pre-operative patients.

The emergence of deep learning analyses now offers the opportunity to capture previously unmeasured risk markers and simultaneously assess complex interactive relationships from readily available clinical resources for risk prediction.[20-22] One such ideal resource for perioperative risk discrimination is the 12-lead electrocardiogram (ECG). ECGs are inexpensive, non-invasive, and rapid diagnostic tests that are routinely obtained in the preoperative setting as per clinical guidelines.[10, 23] Prior studies have applied deep learning analyses of ECG waveforms obtained in other settings to identify clinical traits and outcomes not previously associated with conventional ECG measures or even expert human ECG interpretations.[5,24-28] Therefore, we hypothesized deep learning analyses applied to a single pre-operative ECG could reliably discriminate post-operative mortality and improve upon established clinical approaches to pre-operative assessment. To evaluate this hypothesis, we conducted a comprehensive study of an AI algorithm trained on perioperative ECGs and evaluated the performance on cohorts from three independent healthcare systems.

# METHODS

**<u>Derivation Cohort:</u>**

All patients undergoing inpatient procedures at Cedars-Sinai Medical Center between January 1, 2015 and December 31, 2019 were included in the derivation cohort. During this time,153,465 patients aged 18 years or older underwent 261,328 procedures in the operating room, catheterization laboratory, and endoscopy suite. Of this source population, 45,969 patients who had a complete ECG waveform image available for at least one 12-lead ECG performed within 30 days prior to the procedure date were included in the analysis. These patients contributed 112,794 ECGs preceding 59,975 procedures to the analyses (**Figure 1**).

From this derivation cohort, patients were randomly split 8:1:1 into a subset of 36,839 patients (contributing 90,633 ECGs) for training, 4,549 patients (contributing 11,217 ECGs) for internal validation, and the remaining 4,581 patients (contributing 10,944 ECGs) for final algorithm test analyses. All ECG waveform data were acquired through the clinical enterprise data warehouse at a sampling rate of 500 Hz and extracted as 10 second, 12x5000 matrices of amplitude values. ECGs with missing leads were excluded from analyses. Associated clinical data for each patient were also obtained from the electronic health record. The study was approved by the Cedars-Sinai Medical Center, Stanford University, and Columbia University Institutional Review Boards.

*Clinical and Outcome Assessments*

Patient demographic, clinical, and outcomes data were assessed from the electronic health record at the time of each procedure. From these data, the pre-operative clinical characteristics needed for calculating the revised cardiac risk index (RCRI)[1] were identified, including: coronary artery disease, congestive heart failure, stroke or transient ischemia attack, pre-operative insulin use, creatinine greater than 2mg/dL, and elevated risk surgery as defined by American College of Cardiology and American Heart Association guidelines.[4] For the main analysis outcome was death during the hospitalization or during readmissions within 30 days.

As a secondary analysis, we evaluated the deep learning algorithm's performance in evaluating a composite outcome of MACE that included non-fatal major adverse cardiovascular events, defined by presence of post-operative myocardial infarction, cardiac arrest, heart block, pulmonary edema, as well as mortality.[2] Procedural complications were identified using relevant post-operative diagnoses that were present for the first time after procedure date or at discharge. Outcomes were adjudicated up to 30 days after date of procedure, with multiple procedures having independent outcomes windows based on procedure date. If the same outcome fell within the 30-day window for multiple procedures, the outcome was attributed to each of the procedures. Diagnoses were encoded by International Classification of Diseases (ICD)-9 or ICD-10 codes, and MACE was identified using previously validated criteria from the EHR.[29-31] A random subset of 100 patients with MACE were manually evaluated by chart review. MACE adjudication was correct in 88 (88%) patients, and 81 (81%) of clinician evaluated RCRI scores were the same as the EHR diagnosis code based RCRI score. The mean RCRI score by manual review was 1.46, compared to 1.34 by automated evaluation, a mean difference of 0.13 between diagnoses present in clinical notes but not present in ICD9 codes.

*Electrocardiogram Assessments*

We trained and validated a deep learning algorithm based on waveform signals from a single pre-operative 12-lead ECG, termed PreOpNet, on the outcome of post-operative mortality (**Figure 1**). The input of the model was a 12-lead ECG obtained within 30 days prior to an operative procedure, and the output were the hospitalization level outcomes following that procedure. Patients with multiple procedures were treated independently during model training, with each ECG paired with the most proximal subsequent procedure. To maximize sample size, similar to prior studies of deep learning using electrocardiography,[32] all ECGs in the window of interest were utilized in the training set as training examples. In the internal validation and test cohorts, the analysis was limited to a single ECG most proximal to the procedure of interest for an individual patient to mimic how the model would be applied in clinical practice. When used, clinical features were input into the last fully connected layer prior to model output. Models were trained using the PyTorch deep learning library.

**Deep Learning Algorithm Development**

Based on prior literature regarding lightweight deep learning model architecture design and neural architecture search[33, 34], PreOpNet was designed to analyze 12-lead ECG waveform data starting with atrous convolutions followed by subsequent multi-channel 1D convolutions. The number of layers paralleled the design of EfficientNet[34], and to optimize model runtime and minimize model complexity, the number of layers were limited to less than 1/10th the size of previously described architectures.[5–7] After initial atrous layers, PreOpNet incorporated convolutional layers with an inverted residual structure where the input and output are bottleneck layers with an intermediate expansion layer.[33] In each set of expansion layers with bottleneck layers preceding and succeeding, the number of input channels gradually increased to allow for integration of information across ECG leads. The deep learning model had input only of the 12-lead ECG waveform data, and comparison non-deep learning models included clinical data including age and sex as well as structured ECG information.

The model was initialized with random weights and trained with a loss function of binary cross entropy for 100 epochs using an ADAM optimizer with an initial learning rate between 5e-3 and 1e-4. Early stopping was performed based on validation dataset's area under the receiver operating curve. The atrous convolution's dilation and step size was grid-searched by hyperparameter tuning for optimal AUC with all other hyperparameters held constant (**Supplemental Figure 1**). Local Interpretable Model-agnostic Explanations[35] was used to identify and visualize relevant features in the ECG used for model decision making.

**Statistical Analyses within the Internal Test Cohort:**
**Evaluation of Model Performance and Comparison to an Established Risk Calculator**
Following algorithm development in the training and internal validation datasets, the ability of PreOpNet to discriminate the primary outcome of post-operative mortality in the held-out internal test dataset was assessed using the area under the curve (AUC) of the receiver operating characteristic curve (ROC). To understand the models performance in key patient populations, independent analyses were performed limited to patients undergoing cardiac surgery, non-cardiac surgery, interventional endoscopy suite or catherization laboratory procedures, and patients with known cardiovascular disease or undergoing intermediate or high-risk surgeries.

The performance of PreOpNet were compared in the held-out internal test dataset with an established risk calculator (RCRI score)[8], conventional ECG measures and interpretations, and alternate algorithms based on ECG measurement data.

The association of PreOpNet with perioperative outcomes was compared with that of an RCRI score >2, a commonly used threshold to identify high perioperative risk associated with a MACE rate greater than 8% in meta-analyses.[23,27] In the training set, 15% of the cohort had an RCRI >2; thus a similar threshold for high perioperative risk (top 15%) was set for PreOpNet. Odds ratios, sensitivity, and specificity for post-operative mortality and MACE were estimated at this threshold.[36] Secondary analyses with MACE as the outcome were also performed. The continuous and categorical net reclassification index (NRI) for mortality and MACE rate associated with the addition of PreOpNet to an RCRI score >2 was calculated.[37] 10,000 bootstrapped samples were used to obtain 95th confidence intervals for each estimate. To discern the potentially most informative features of the ECG waveform, in the context of comparing performance to that of the RCRI, 0.5% of the waveform for 1000 samples per study were iteratively randomly perturbed to identify which changes most impacted model performance.[35] Implementation timings were evaluated using Python's timeit module.

***Comparison to Conventional and Alternate ECG Measures***

PreOpNet performance was compared with that of a non-deep learning model (eXtreme Gradient Boost (XGB)) which considers the clinical risk factors used in the RCRI score (history of ischemic heart disease, congestive heart failure, cerebrovascular disease, insulin use, preoperative creatinine greater than 2 mg/dL, or elevated risk procedure). The PreOpNet algorithm performance was also compared with that of models trained on the human interpretable features including: (i) clinical variables from the RCRI score; (ii) age and all ECG measurements obtained from MUSE including heart rate, axis, and intervals, and physician ECG interpretations; and, (iii) the combination of age, ECG measurements, physician ECG interpretations, and clinical variables. Physician interpretations in the clinical over-read section of MUSE were coded as distinct categorial variables and considered as potential predictors in comparison models (**Supplemental Table 5**).

**External Validation of PreOpNet for Discrimination of Post-Operative Mortality**

To assess algorithm performance in other hospital settings, the PreOpNet algorithm was applied without any additional further fine tuning or training to patients from two separate external healthcare systems. To maximize rigor of external validation, patient data was not transferred across institutions: rather, investigators from each external institution independently collected outcomes information and ran model inference on the their datasets to report summary statistics. The Stanford Healthcare (SHC) cohort included 101,375 patients contributing 162,540 pre-operative ECGs from May 1, 2007 to June 30, 2018. All clinical characteristics and outcomes data were provided through the Stanford Research Data Repository Observational Medical Outcomes Partnership (OMOP) common data model. ECG waveform data was provided through the TraceMaster (Philips Healthcare) data management system and preprocessed with a low pass filter to further correct for wandering baselines and normalization of waveform data. The Columbia University Medical Center (CUMC) cohort included 9,028 patients contributing 9,028 pre-operative ECGs from January 1, 2020 to March 31, 2020. At CUMC, clinical characteristics and outcomes data were obtained from the clinical enterprise data warehouse and ECG waveform data was obtained from Muse (GE Healthcare) data management system. For each of the two external cohorts, the AUC for post-operative mortality from analyses of a single pre-operative ECG was calculated. Procedures were linked to the most proximal preceding ECG performed within 30 days prior and post-operative mortality was assessed as mortality within 30 days after the procedure or during the hospitalization. The most common procedures are summarized in Supplementary Table 8.

All code and analytical methods applied for the algorithm and analyses are available at https://github.com/ecg-net/PreOpNet. The patient data is not available given the potentially identifiable nature of the associated data.

RESULTS

**Derivation (CSMC) Cohort Characteristics**

The CSMC cohort included 45,969 patients who underwent 59,975 inpatient procedures between years 2013 and 2019 (**Figure 1**). The mean age at the time of pre-operative ECG was 65.1±15.9 years, 45% were women, and 22% had pre-existing coronary artery disease as shown in **Table 1**. There were 1,065 (1.8%) subsequent deaths and 1,730 (2.9%) post-procedural MACE events during the hospitalization in the cohort. Clinical characteristics and index hospitalization outcomes are also shown in **Table 1** for the training, internal validation, and internal test subsets of the cohort. Compared with those patients who were excluded based on not having ECG data within 30 days of their procedure, patients with pre-operative ECGs were more likely to be older, male, and with more cardiovascular risk factors (**Supplementary Table 7**).

**Discrimination of Post-Procedural Mortality at CSMC**

For the outcome of mortality, the PreOpNet algorithm was developed in the derivation cohort using training, validation, and test datasets. In the held-out test dataset, the algorithm was then shown to discriminate mortality with an AUC of 0.83 (95% CI 0.79-0.87) (**Table 2**). By contrast, the conventional RCRI score discriminated post-operative mortality with an AUC of 0.67 (0.61-0.72). In the 15% of patients who had an RCRI score of 2 or greater, the unadjusted odds ratio for post-operative mortality was of 2.08 (95% CI, 0.77-3.50). In comparison, patients in the top 15% of the PreOpNet algorithm had an adjusted odds ratio of 9.17 (95% CI, 5.85-13.82) for post-operative mortality. The addition of the components of the RCRI score to the PreOpNet algorithm trained only on 12-lead ECG waveforms did not significantly improve model performance (AUC 0.83; 95% CI, 0.77-0.89) in the test dataset. There were no significant differences in model performance in patient subsets by age, sex, or race (Supplemental Table 5). At the pre-specified calibrated threshold of risk comparable to a RCRI score ≥ 2, the PreOpNet algorithm demonstrated a specificity of 0.87 (0.86-0.88) and sensitivity of 0.57 (0.48-0.68) for post-operative mortality. In comparison, the RCRI score ≥ 2 had slightly higher specificity of 0.94 (0.93-0.94) but much lower sensitivity of 0.12 (0.05-0.19). At this threshold, the positive predictive value was slightly higher for PreOpNet as compared to the RCRI score. (**Table 2**).

**Discrimination of Post-Procedural Mortality in Important Clinical Scenarios**

PreOpNet performed well both in patients undergoing major surgical procedures in the operating room as well as patients undergoing procedures in the catherization laboratory or endoscopy suite. For patients with surgeries in the operating room, PreOpNet discriminated post-operative mortality with an AUC of 0.84 (0.76 – 0.92), compared to an AUC of 0.70 (0.61 – 0.78) for the RCRI score. For patients with procedures in the catherization lab or endoscopy suite, PreOpNet discriminated post-operative mortality with an AUC of 0.83 (0.78 – 0.98), compared to an AUC of 0.66 (0.60 – 0.72) for the RCRI score. PreOpNet performed similarly in discriminating mortality in patients undergoing cardiovascular surgery with an AUC of 0.85 (0.77 – 0.92) and patients undergoing non-cardiac surgery with an AUC of 0.83 (0.79 – 0.88). In patients undergoing cardiac surgery, the RCRI score discriminated post-operative mortality with an AUC of 0.62 (0.52-0.72), and in patients undergoing non-cardiac surgery, the RCRI score discriminated post-operative mortality with an AUC of 0.70 (0.63-0.77).

Given that ECGs are often not obtained in low-risk patients undergoing low-risk procedures, a secondary analysis was performed in patients most likely to be considered at least moderate-risk

– patients either with known cardiovascular disease or those undergoing elective intermediate-risk or high-risk surgery. Without additional subset-specific fine-tuning, the PreOpNet algorithm discriminated post-operative mortality in this subset with an AUC of 0.80 (0.71-0.88). In clinical practice, pre-operative risk assessment most commonly occurs in the elective procedural setting. Thus, secondary analyses were performed limited to those patients in the CSMC test cohort who were undergoing elective procedures (3,691 patients contributing 5,165 ECGs). In this setting, the PreOpNet algorithm discriminated post-operative mortality with an AUC of 0.80 (0.67-0.92).

**Comparison of PreOpNet Performance with Alternative ECG Assessment Models**

The AUC for post-operative mortality was greater for PreOpNet as compared with traditional ECG measures (e.g. heart rate, axis, and intervals), physician ECG interpretations, or clinical variables in the hold-out test dataset (**Supplementary Table 2**). An XGB model on RCRI variables performed less well than PreOpNet and performed similarly to the RCRI score in discriminating post-operative mortality with an AUC of 0.70 (0.66 – 0.74). To clarify how functionality of the PreOpNet algorithm might be interpreted in clinical context, Local Interpretable Model-agnostic Explanations (LIME)[9,10] was applied to the data. The LIME sensitivity analyses highlighted features of the QRS complexes as the most relevant to model decision making; in addition, precordial premature contractions and intraventricular block on the precordial leads were also frequently highlighted as part of this secondary analysis (**Figure 3**).

**Discrimination of Post-Procedural Mortality in External Validation Cohorts**

To assess external validity of the PreOpNet algorithm, the discriminatory ability was evaluated in two external health system cohorts. The external test evaluation cohorts included 101,375 patients in the Stanford Healthcare (SHC) system contributing 162,540 ECGs and 9,028 patients in the Columbia University Medical Center (CUMC) system contributing 9,028 ECGs. In the SHC cohort, the post-operative mortality rate was 1.3% and PreOpNet discriminated this outcome with an AUC of 0.75 (0.74-0.76). In the CUMC cohort, the post-operative mortality rate was 1.6% and the algorithm discriminated this outcome with an AUC of 0.79 (0.75-0.83) (**Table 2**). The PreOpNet algorithm pre-specified high risk group (>15%) had an unadjusted odds ratio of 5.88 (5.00-7.00) in SHC and 6.20 (3.87-10.41) in CUMC for mortality. Results from analyses of specificity, sensitivity, and positive and negative predictive value were similar in the external validation cohorts when compared with results observed in the CSMC cohort (**Table 2**).

**Major Cardiovascular Events (MACE) within the Internal Test Dataset.**

Secondary analyses were performed within the internal test dataset that utilized a combined secondary MACE outcome that included non-fatal major cardiovascular events and post-operative mortality. For this secondary outcome, the PreOpNet algorithm discriminated events in the held-out internal test dataset with an AUC of 0.77 (0.73-0.80) whereas the RCRI score had an AUC of 0.63 (0.59-0.68). Patients with an RCRI score of 2 or greater had an unadjusted odds ratio of 1.67 (0.77-2.68) for MACE when compared with those with an RCRI score <2. By contrast, patients identified by the PreOpNet algorithm to be high risk had an unadjusted odds ratio 5.38 (3.75-7.49) for MACE (**Table 2**). The PreOpNet algorithm demonstrated a specificity of 0.88 (0.88-0.89) and sensitivity of 0.41 (0.33-0.49) while the RCRI score again had higher specificity at 0.94 (0.93-0.94) but much lower sensitivity at 0.10 (0.05-0.15). The PPV for MACE was higher for PreOpNet as compared with the RCRI with similar negative predictive values (**Table 2**).

**Clinical Risk Assessment and Reclassification for MACE**

The ability of the PreOpNet algorithm to reclassify risk in the internal hold-out test dataset was also evaluated. When compared with the RCRI score, application of the PreOpNet algorithm led to significant improvement in the continuous net reclassification index (NRI 0.53; 95% CI, 0.38 to 0.68). In categorical analyses using the pre-specified threshold of risk (top 15th percentile PreOpNet prediction), 981 (82.4%) of 1190 patients originally classified as high-risk for MACE by the RCRI score (RCRI ≥2) were identified as low-risk by PreOpNet (Supplemental Table 3). Of these high-to-low reclassified risk patients, 33 (3.4%) patients experienced MACE. By contrast, of the 4,739 patients classified as low-risk by RCRI (RCRI <2), the PreOpNet algorithm reclassified 327 (6.9%) to be high risk; of these patients, 26 (8.0%) patients experienced MACE. Despite a fair amount of reclassification, the categorical NRI at this cut-point was not significant (NRI 0.06; 95% CI -0.04 to 0.18) for MACE.

**Technical Application**

To facilitate the potential pragmatic application of the PreOpNet algorithm as a decision support tool in the clinical setting, the algorithm was optimized to run on a standard-build clinical workstation without graphical processing units (GPUs) and also developed a web application to facilitate access by clinicians (**Supplemental Figure 4**). In a series of repeated runtime experiments, the PreOpNet application accessed on a standard clinical workstation (Windows 10 64-bit OS, 3.8 GHz processor) was able to intake image data from 50 de novo ECGs and output post-operative risk estimates within 0.032±0.004 seconds per ECG for the local software installation and 0.041±0.006 seconds per ECG for the web application accessed by a mobile phone.

**DISCUSSION**

In a large cohort of patients undergoing inpatient procedures, a deep learning algorithm utilizing the waveforms of a single pre-operative 12-lead ECG identified risk for post-operative death for cardiac surgeries, non-cardiac surgeries, and catherization lab interventions. Compared with a widely used standard peri-operative risk assessment tool and alternative ECG assessment tools, PreOpNet was able to more effectively identify high-risk patients who went on to experience postoperative mortality. Further, the accuracy of PreOpNet for discriminating post-operative mortality was re-affirmed in two external healthcare system cohorts with diverse patient populations. Additionally, in secondary analyses, PreOpNet identified high-risk patients who went on to experience MACE within the internal test dataset in the CSMC patient population. To our knowledge, PreOpNet is the first deep learning architecture designed to aid clinicians in discriminating post-operative outcomes.

The current study builds on prior efforts to improve efficacy as well as efficiency of the pre-operative risk assessment. Conventional approaches have evolved over the past three decades to include clinical risk factors as well as both biomarkers and imaging diagnostics.[2, 7-10] However, it has proved difficult to achieve greater than modest accuracy in the prediction of post-operative outcomes.[4-6 6, 11-14] At the same time, concerns surrounding the costs and burden of potentially unnecessary pre-operative testing have continued to mount.[38, 39] Many experts have recognized that a persistent challenge in achieving more accurate prediction of post-operative outcomes is the marked heterogeneity of patients at risk, which amplifies the potential importance of unmeasured variables.[2,3,5] To address this challenge, algorithm development focused on deriving new prognostic information from the pre-operative 12-lead ECG given its wide accessibility as a diagnostic test frequently ordered in clinical practice.[23]

Over the last several years, the 12-lead ECG has been the subject of deep learning algorithm development for potential clinical applications, with promising results. Previously applied to improve detection of occult cardiovascular traits (e.g. atrial fibrillation,[32] hypertrophic cardiomyopathy[40], cardiac amyloidosis[41], and ventricular dysfunction[42, 43]), recent deep learning ECG models have been shown to also identify non-cardiovascular specific traits such as liver disease,[11] anemia,[27] age,[28] and long-term mortality.[24, 44] These latter results highlight the potential of deep learning methods to extract broad as well as specific novel information from ECG waveforms. Therefore, to optimize performance for the task of pre-operative risk assessment, our algorithm used readily available ECG waveform data collected from large diverse real-world patient cohorts. The resulting PreOpNet algorithm demonstrated ability not only to discriminate post-operative mortality, but to do so while outperforming a conventional clinical risk score and human-interpretable features from the ECG. As indicated in other clinical fields where artificial intelligence has been shown to leverage latent features from a medical image to refine diagnosis or prognostication,[22, 45, 46] our results indicate the potential value of the PreOpNet algorithm to augment clinical decision making for pre-operative risk assessment.

For deep learning algorithms to be readily integrated in the clinical practice setting, implementation needs to pose little to no additional burden to existing clinical workflows. Although not the primary focus of the current study, we took several steps towards assessing the potential application of PreOpNet in practice. Recognizing that many image-based, including

ECG focused algorithms, tend to be computationally resource intensive,[25, 47] the PreOpNet algorithm was optimized to run on a standard-build clinical workstation that does not include graphical processing units (GPUs) as part of its configuration. The runtime from local and web-based installations was assessed and found to have an acceptable real-time implementation duration of <1 second. While additional studies would be needed to fully evaluate implementation efficacy, the opportunity to avoid need for any manual data entry – which is the requirement for most currently available risk score calculators – offers a promising option for integration into existing workflows.

Several limitations of this preliminary study merit consideration. Many ambulatory procedures for clinically assessed low-risk patients do not involved acquiring a pre-operative ECG. Therefore, this algorithm may not be applicable to such lower risk patients. The non-fatal MACE outcomes for the secondary analysis were derived from the EHR, which has been shown to under-report post-procedural complications compared to independent adjudication[48,12,13]. Although a random sample of events for non-fatal MACE outcome were manually adjudicated, the sensitivity of the EHR method for MACE adjudication is unknown. To further minimize data heterogeneity when assessing cardiovascular outcomes from the EHR, external validation analyses were limited to the primary outcome of mortality. Also, all analyses were performed on retrospective cohorts. Therefore, additional prospective validation studies are needed in large and diverse external cohorts – particularly of the exploratory secondary MACE endpoint– to precisely evaluate PreOpNet's performance in discriminating events.[43, 49] Notwithstanding these limitations, the current study offers several strengths including the ability to leverage internal training, validation, and test datasets within a large derivation cohort of patients undergoing inpatient procedures over a decade of time. The algorithm was also able to be externally validated for post-procedural mortality in two large, diverse medical centers.

In summary, our findings demonstrate how a novel deep learning algorithm, applied to a single pre-operative ECG, can improve discrimination of post-operative adverse outcomes while running efficiently on a standard clinical workstation. Recognizing that clinicians have limited time for making clinical assessments and decisions around potential post-procedural outcomes, conventional risk calculators using easily accessible information have been recommended by professional society practice guidelines to aid in peri-operative risk stratification.[10, 23] The opportunity to implement potentially more informative and easier to use prediction algorithms, in a manner that integrates with existing clinical workflows, offers a potential path towards improving post-operative outcomes. These promising results warrant further studies to determine the prospective validity of deep learning algorithms for prognosticating post-procedural risk.

**Acknowledgements**

D.O. is supported by NIH K99 HL157421-01. J.W.H. and B.H. is supported by the NSF Graduate Research Fellowship. NY is supported by NIH 5T32HL116273-07. J.H.C. is supported by NIH R56LM013365, NSF SPO 181514, and Gordon and Betty Moore Foundation Grant GBMF8040. J.Y.Z. is supported by NSF CCF 1763191, NIH R21 MD012867-01, NIH P30AG059307 and by a Chan-Zuckerberg Biohub Fellowship. NRC and CMA are supported by HL 091069 and HL11669.

**Author Contributions**

DO, JWH, NS, JT, JE, RKS, NY, PB, EP, JHC, JE, MT retrieved and quality controlled all ECG data and merged electronic medical record data. JWH, PE, TP, JET, JHC, JT, AP, EAA, JYZ were responsible for external validation. DO, JT, JWH, BH, GD, JTS, BC, PE, NC, SC developed and trained the deep learning algorithms, performed statistical tests, and created all the figures. DO, MP, NY, MN, SSC, EAA, SC, CMA performed clinical evaluation of model performance. DO, NS, JT, SC, CMA wrote the manuscript with critical review and feedback by all authors.

**Disclosures**

The authors report nothing to disclose.

**FIGURE LEGENDS**

**Figure 1. CSMC site cohort sampling.** A. ECGs within 30 days of an inpatient procedure was selected for the study and paired with post-operative outcomes. B. A novel light-weight model architecture was trained to discriminate post-operative mortality and complications, with input of the nearest 12-lead ECG. C. Consort Diagram. ECG = electrocardiogram. MACE = Major adverse cardiovascular events. CSMC = Cedars-Sinai Medical Center, SHC = Stanford Healthcare, and CUMC = Columbia University Medical Center.

**Figure 2. PreOpNet workflow and results.** Performance of PreOpNet at Cedars-Sinai Medical Center (CSMC, red), Stanford Healthcare (SHC, green), and Columbia University Medical Center (CUMC, blue), and Revised Cardiac Risk Index in discriminating (C) post-operative mortality and (D) major adverse cardiovascular events. E. Rate of mortality and complications by percentile of risk identified by PreOpNet. ECG = electrocardiogram. RCRI = Revised Cardiovascular Risk Index.

**Figure 3. Interpretability Analysis of PreOpNet.** Select electrocardiograms before procedures with positive and negative outcomes highlighting most relevant features as determined by interpretability analysis. A) Discrimination of Mortality, B) Discrimination of Major Adverse Cardiovascular Events.

**Table 1. Characteristics and Outcomes of the Derivation Cohort**

|  | Total Cohort | Training Subcohort | Validation Subcohort | Test Subcohort |
|---|---|---|---|---|
| No. of Patients | 45,969 | 36,839 | 4,549 | 4,581 |
| No. of Procedures | 59,975 | 48,033 | 6,013 | 5,929 |
| No. of ECGs | 112,794 | 90,633 | 11,217 | 10,944 |
|  |  |  |  |  |
| **Demographic and Clinical Characteristics** |  |  |  |  |
| Age, years (SD) | 65.1 (15.9) | 65.2 (15.8) | 65.0 (16.4) | 64.6 (16.1) |
| Female, n (%) | 27091 (45.2) | 21744 (45.3) | 2706 (45.0) | 2641 (44.6) |
| Heart Failure, n (%) | 11000 (18.3) | 8689 (18.1) | 1192 (19.8) | 1119 (18.9) |
| Diabetes Mellitus, n (%) | 9939 (16.6) | 7880 (16.4) | 998 (16.6) | 1061 (17.9) |
| Hypertension, n (%) | 20888 (34.8) | 16638 (34.6) | 2187 (36.4) | 2063 (34.8) |
| Coronary Artery Disease, n (%) | 13019 (21.7) | 10421 (21.7) | 1293 (21.5) | 1305 (22.0) |
| Stroke, n (%) | 3392 (5.7) | 2697 (5.6) | 352 (5.9) | 343 (5.8) |
| Renal Disease, n (%) | 5683 (9.5) | 4475 (9.3) | 640 (10.6) | 568 (9.6) |
|  |  |  |  |  |
| **Procedure Types** |  |  |  |  |
| Cardiovascular, n (%) | 24882 (41.5) | 19840 (41.3) | 2595 (43.2) | 2447 (41.3) |
| Intraperitoneal; intrathoracic; suprainguinal vascular, n (%) | 10843 (18.1) | 8679 (18.1) | 1074 (17.9) | 1090 (18.4) |
| Insulin use prior to admission, n (%) | 3762 (6.3) | 3022 (6.3) | 366 (6.1) | 374 (6.3) |
| Creatinine > 2 mg/dL, n (%) | 5602 (9.3) | 4408 (9.2) | 637 (10.6) | 557 (9.4) |
| RCRI > 2, n (%) | 3489 (5.8) | 2784 (5.8) | 327 (5.4) | 378 (6.4) |
|  |  |  |  |  |
| **Post-Operative Outcomes** |  |  |  |  |
| Death during admission, n (%) | 1065 (1.8) | 865 (1.8) | 109 (1.8) | 91 (1.5) |
| Cardiovascular death, n (%) | 32 (0.1) | 21 (0.0) | 7 (0.1) | 4 (0.1) |
| Major cardiovascular events, n (%) | 1730 (2.9) | 1400 (2.9) | 190 (3.2) | 140 (2.4) |
| Cardiac arrest, n (%) | 227 (0.4) | 182 (0.4) | 27 (0.4) | 18 (0.3) |
| Myocardial infarction, n (%) | 253 (0.4) | 203 (0.4) | 30 (0.5) | 20 (0.3) |
| Heart block, n (%) | 160 (0.3) | 129 (0.3) | 22 (0.4) | 9 (0.2) |
| Pulmonary edema, n (%) | 48 (0.1) | 40 (0.1) | 4 (0.1) | 4 (0.1) |

ECG, electrocardiogram

**Table 2. Discrimination of Post-Operative Outcomes by the PreOpNet or RCRI Algorithms.**

| Outcome | Model (Cohort) | AUC | Specificity | Sensitivity | PPV | NPV | Odds Ratio* |
|---|---|---|---|---|---|---|---|
| Death | PreOpNet (CSMC) | 0.83 (0.79-0.87) | 0.87 (0.86-0.88) | 0.57 (0.48-0.68) | 0.06 (0.05-0.08) | 0.99 (0.99-0.99) | 9.17 (5.85-13.82) |
| | PreOpNet (CUMC) | 0.79 (0.75-0.83) | 0.86 (0.80-0.92) | 0.49 (0.48-0.50) | 0.03 (0.02-0.03) | 1.00 (0.99-1.00) | 6.20 (3.87-10.41) |
| | PreOpNet (SHC) | 0.75 (0.74-0.76) | 0.93 (0.91-0.94) | 0.32 (0.32-0.32) | 0.02 (0.02-0.02) | 1.00 (1.00-1.00) | 5.88 (5.00-7.00) |
| | RCRI | 0.67 (0.61-0.72) | 0.94 (0.93-0.94) | 0.12 (0.05-0.19) | 0.03 (0.01-0.05) | 0.99 (0.98-0.99) | 2.08 (0.77-3.50) |
| MACE | PreOpNet (CSMC) | 0.77 (0.73-0.80) | 0.88 (0.88-0.89) | 0.41 (0.33-0.49) | 0.08 (0.06-0.10) | 0.98 (0.98-0.99) | 5.38 (3.75-7.49) |
| | RCRI | 0.63 (0.59-0.68) | 0.94 (0.93-0.94) | 0.10 (0.05-0.15) | 0.04 (0.02-0.06) | 0.98 (0.97-0.98) | 1.67 (0.77-2.68) |

AUC = Area under the curve. MACE = Major adverse cardiovascular events. RCRI = Revised Cardiovascular Risk Index. CSMC denotes results for the internal held-out test dataset from the derivation cohort, and SHC and CUMC denote results for the external validation cohorts.

*Unadjusted odds ratios for the outcome of interest (death or MACE) were calculated for patients in the ≥85th percentile of risk compared to those in the <85th percentile of risk, as determined by either PreOpNet or the RCRI (corresponding go an RCRI score ≥2).

**Figure 1.**

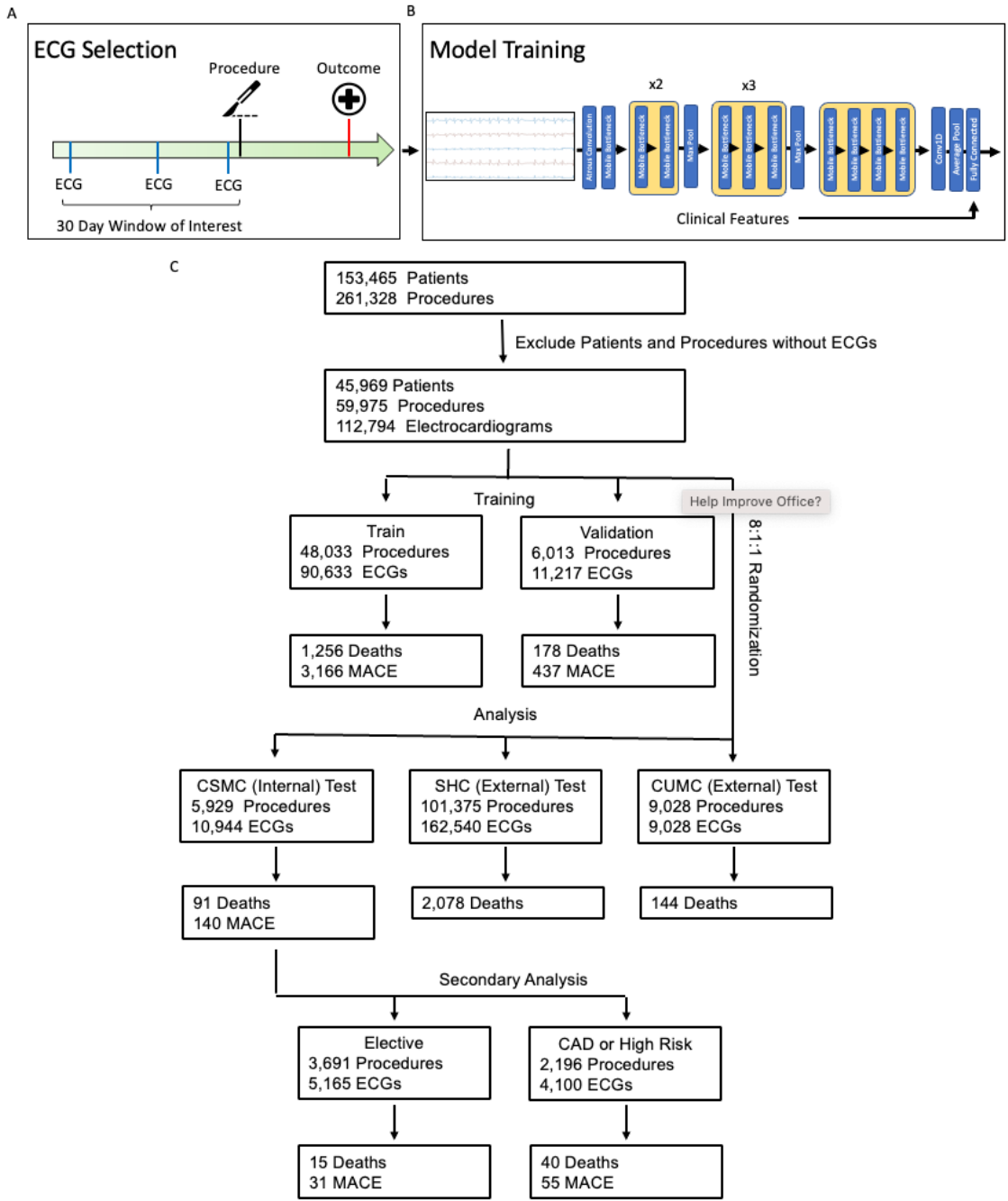

**Figure 2.**

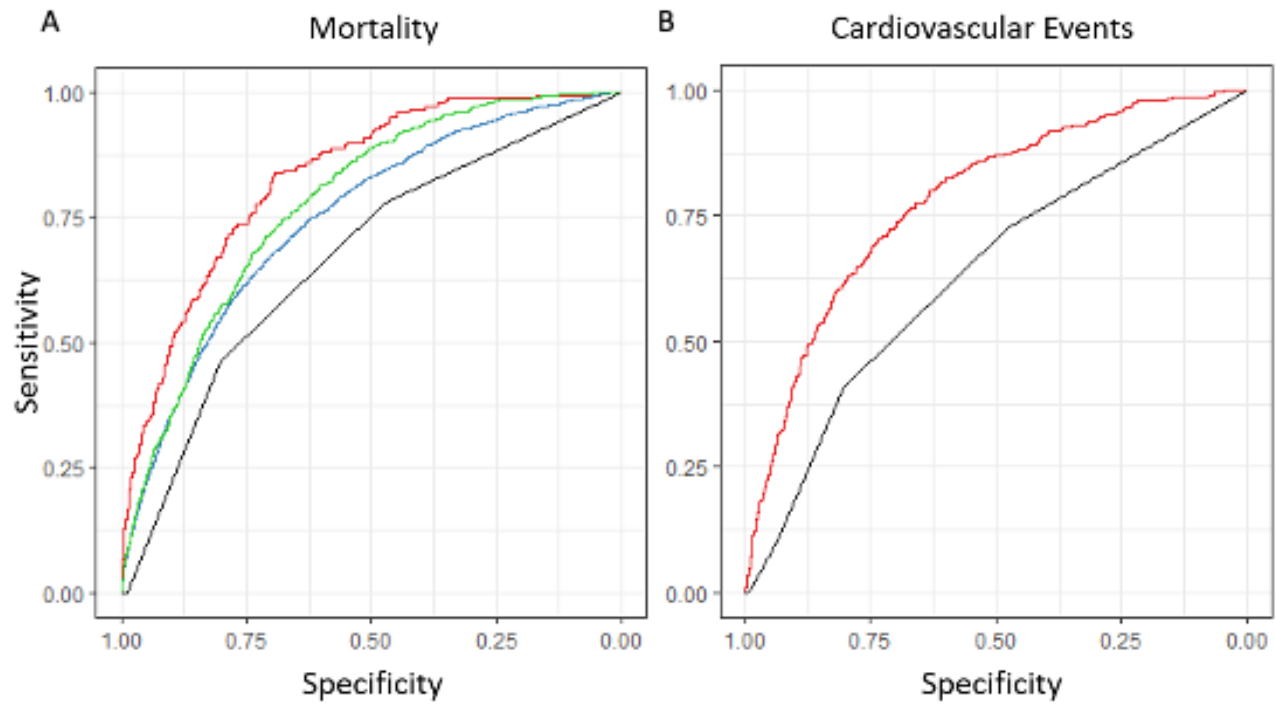

**Figure 3.**

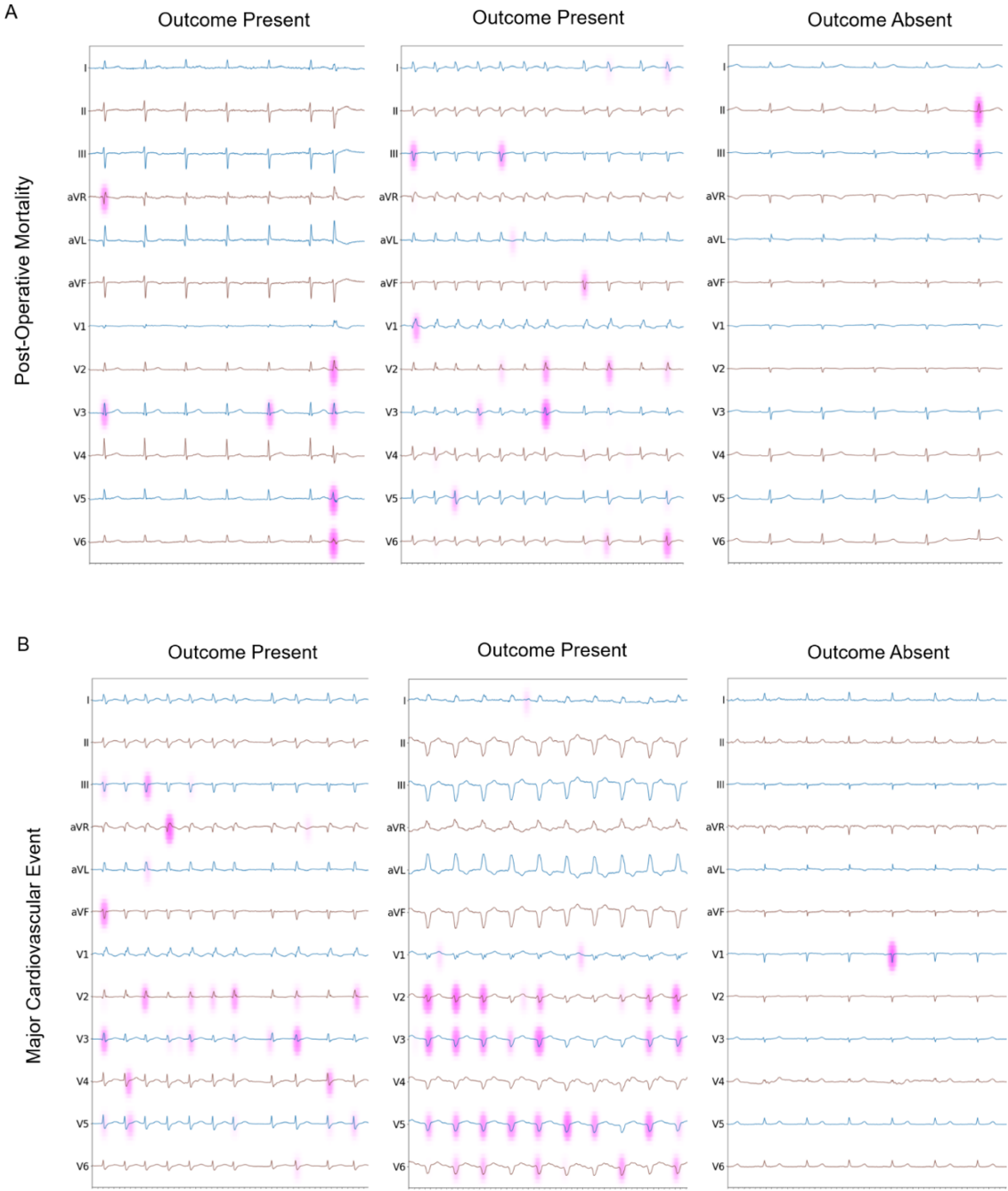

# SUPPLEMENTAL MATERIAL

**Supplemental Figure 1: Hyperparameter Sweep**

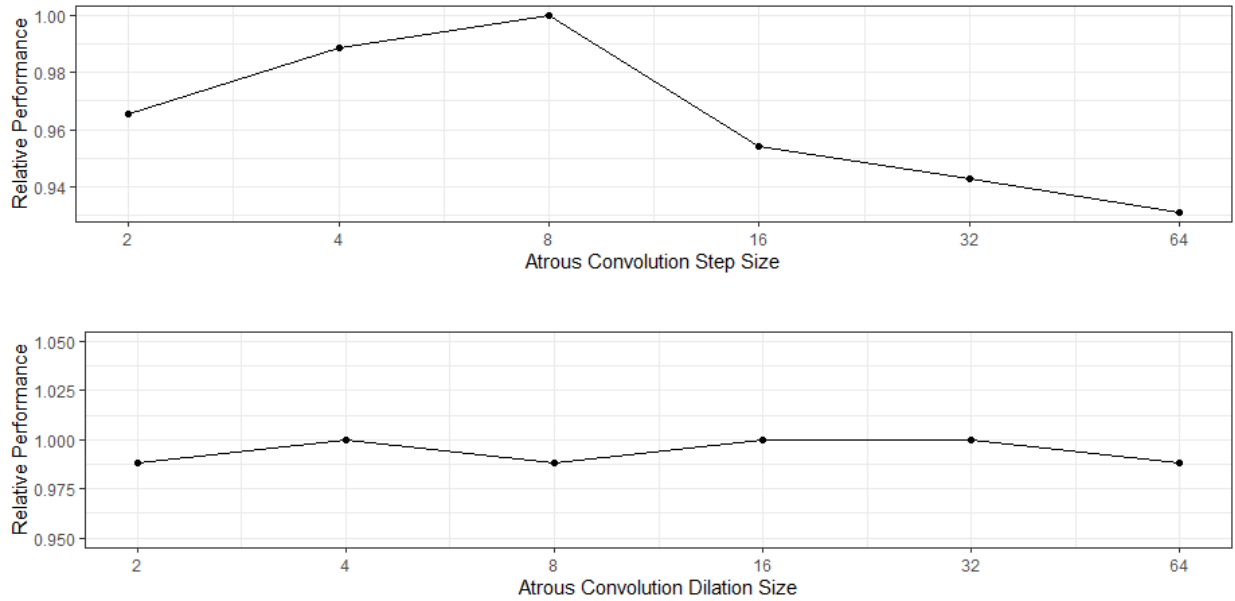

**Supplemental Figure 2. Web Application Deployment**

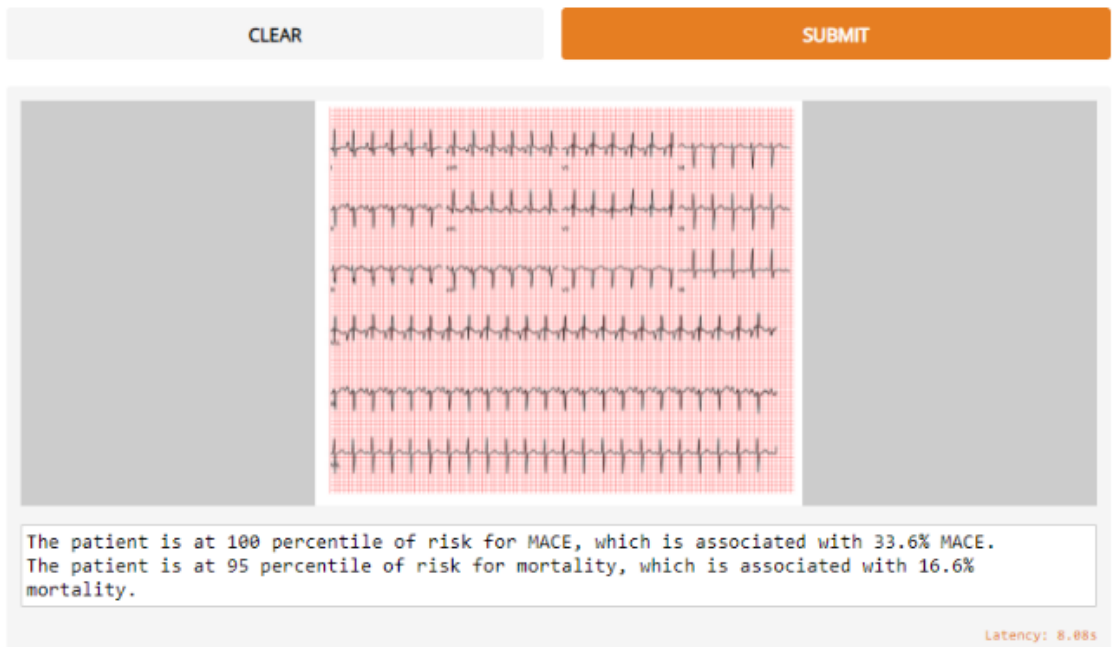

**Supplemental Table 1: PreOpNet model performance at various thresholds of risk**

| Outcome | Threshold | Odds Ratio For Event | Specificity | Sensitivity | PPV | NPV |
|---|---|---|---|---|---|---|
| Death | Max Youden's index | 9.48 (5.99-14.91) | 0.82 (0.81-0.83) | 0.66 (0.56-0.76) | 0.06 (0.04-0.07) | 0.99 (0.99-1.00) |
| | Top 7% of Risk | 8.83 (5.57-13.20) | 0.90 (0.89-0.91) | 0.48 (0.38-0.59) | 0.07 (0.05-0.09) | 0.99 (0.99-0.99) |
| | Top 15% of Risk | 9.17 (5.85-13.82) | 0.87 (0.86-0.88) | 0.57 (0.48-0.68) | 0.06 (0.05-0.08) | 0.99 (0.99-0.99) |
| | Top 50% of Risk | 10.47 (5.34-24.47) | 0.53 (0.52-0.54) | 0.89 (0.82-0.96) | 0.03 (0.02-0.03) | 1.00 (0.99-1.00) |
| Death or non-fatal MACE | Max Youden's index | 5.31 (3.41-8.41) | 0.54 (0.52-0.55) | 0.81 (0.75-0.88) | 0.04 (0.03-0.05) | 0.99 (0.99-0.99) |
| | Top 7% of Risk | 6.15 (4.13-8.58) | 0.92 (0.91-0.92) | 0.36 (0.28-0.44) | 0.09 (0.07-0.12) | 0.98 (0.98-0.99) |
| | Top 15% of Risk | 5.38 (3.75-7.49) | 0.88 (0.88-0.89) | 0.41 (0.33-0.49) | 0.08 (0.06-0.10) | 0.98 (0.98-0.99) |
| | Top 50% of Risk | 5.30 (3.40-8.39) | 0.54 (0.52-0.55) | 0.81 (0.75-0.88) | 0.04 (0.03-0.05) | 0.99 (0.99-0.99) |

**Supplemental Table 2. PreOpNet model performance compared to non-deep learning based risk prediction models at CSMC.** The alternative models with inputs of 1) the combination of ECG measurements, physician ECG interpretations, and clinical variables, 2) all ECG measurements obtained from MUSE including heart rate, axis, and intervals, 3) physician ECG interpretations, and 4) clinical variables from the RCRI score.

| Outcome | Model | AUC |
| --- | --- | --- |
| Mortality | PreOpNet | 0.83 (0.78-0.88) |
| | ECG Measurements, Interpretation, RCRI clinical variables | 0.64 (0.63-0.65) |
| | ECG Measurements (Intervals, Axis, HR, etc) | 0.66 (0.65-0.67) |
| | Physician Overread Interpretation of ECGs | 0.64 (0.63-0.65) |
| | Clinical Variables from RCRI Score | 0.65 (0.55-0.74) |
| MACE | PreOpNet | 0.77 (0.74-0.80) |
| | ECG Measurements, Interpretation, RCRI clinical variables | 0.74 (0.73-0.75) |
| | ECG Measurements (Intervals, Axis, HR, etc) | 0.60 (0.59-0.61) |
| | Physician Overread Interpretation of ECGs | 0.63 (0.62-0.63) |
| | Clinical Variables from RCRI Score | 0.61 (0.51-0.70) |

**Supplemental Table 3. Performance and computational complexity of deep learning models.** FLOPs = Floating point operations

|  | FLOPs* | Training Time |
|---|---|---|
| PreOpNet | 57,913,936 | 82 minutes |
| Attia et al. | 95,773,024 | 103 minutes |
| Kwon et al. | 199,355, 858 | 194 minutes |
| Goto et al. | 3,155,739,318 | 1320 minutes |

**Supplemental Table 4. Net Reclassification Table for Major Averse Cardiovascular Events (MACE).** RCRI = Revised Cardiac Risk Index

|  |  | PreOpNet Risk Classification | | Total | No. (%) Reclassified |
|---|---|---|---|---|---|
|  |  | 0-8% Risk of MACE | Greater than 8% Risk of MACE | | |
| **RCRI Score** | **Low Risk (Score < 2)** | | | | |
|  | Number of participants | 4412 | 327 | 4739 | 327 (6.9%) |
|  | Percent of participants | 74.4 | 5.5 | | |
|  | MACE | 57 (1.3%) | 26 (8.0%) | | |
|  | **High Risk (Score >= 2)** | | | | |
|  | Number of participants | 981 | 209 | 1190 | 981 (82.4%) |
|  | Percent of participants | 16.6 | 3.5 | | |
|  | MACE | 33 (3.4%) | 24 (11.5%) | | |
|  | | | | | |
|  | Total | 5393 | 536 | 5929 | 1308 (22.1%) |

**Supplemental Table 5. ECG Measurements and Interpretations used in non-deep learning models.** We trained the non-deep learning models with ECG measurements obtained from MUSE including heart rate, axis, and intervals (column 1) and physician ECG interpretations from the clinical overread section of MUSE as categorical variables (column 2).

| Muse Measurements | Common Physician Interpretations |
|---|---|
| AtrialRate | Normal sinus rhythm |
| PAxis | Prolonged QT interval |
| POffset | Left axis deviation |
| POnset | Inferior infarct |
| PRInterval | Right bundle branch block |
| PharmaPPinterval | 1st degree A-V block |
| PharmaRRinterval | Suggest repeat electrocardiogram |
| QOffset | Nonspecific T wave abnormality |
| QOnset | Left atrial enlargement |
| QRSCount | Premature ventricular complexes |
| QRSDuration | Otherwise normal ECG |
| QTCorrected | No significant change was found |
| QTInterval | Low voltage QRS |
| QTcFrederica | Premature atrial complexes |
| RAxis | T wave abnormality, consider lateral ischemia |
| TAxis | Left Ventricular Hypertrophy |
| TOffset | Borderline ECG |
| VentricularRate | Anterior infarct |
| HeightCM | Septal infarct |
| WeightKG | Left anterior fascicular block |
| PatientAge | Nonspecific ST and T wave abnormality |

**Supplemental Table 6. Performance of PreOpNet stratified by demographic groups**

| Outcome | Demographic | AUC |
|---|---|---|
| MACE | Female | 0.80 (0.74 – 0.87) |
| | Male | 0.74 (0.69 – 0.79) |
| | Black | 0.72 (0.57 – 0.86) |
| | Caucasian | 0.76 (0.71 – 0.81) |
| | Other Ethnicity | 0.82 (0.75 – 0.89) |
| | Under 55 years old | 0.79 (0.68 – 0.89) |
| | 55-75 years old | 0.77 (0.72 – 0.82) |
| | Older than 75 years old | 0.74 (0.66 – 0.81) |
| Death | Female | 0.87 (0.81 – 0.94) |
| | Male | 0.80 (0.75 – 0.85) |
| | Black | 0.84 (0.77 – 0.91) |
| | Caucasian | 0.83 (0.77 – 0.88) |
| | Other Ethnicity | 0.83 (0.74 – 092) |
| | Under 55 years old | 0.84 (0.75 – 0.92) |
| | 55-75 years old | 0.87 (0.83 – 0.91) |
| | Older than 75 years old | 0.75 (0.64 – 0.85) |

**Supplemental Table 7. Characteristics and outcomes of the derivation cohort compared to patients undergoing procedures**

|  | All Procedures | Procedures with ECGs | Training | Validation | Test |
|---|---|---|---|---|---|
| Number of Patients | 153,465 | 45,969 | 36,839 | 4,549 | 4,581 |
| Number of Procedures | 261,328 | 59,975 | 48,033 | 6,013 | 5,929 |
| Number of ECGs | 112,794 | 112,794 | 90,633 | 11,217 | 10,944 |
| **Demographics** | | | | | |
| Age, years (SD) | 58.5 ( 19.0 ) | 65.1 ( 15.9 ) | 65.2 ( 15.8 ) | 65.0 ( 16.4 ) | 64.6 ( 16.1 ) |
| Female, n (%) | 142391 ( 54.5 %) | 27091 ( 45.2 %) | 21744 ( 45.3 %) | 2706 ( 45.0 %) | 2641 ( 44.6 %) |
| Heart Failure, n (%) | 20777 ( 8.0 %) | 11000 ( 18.3 %) | 8689 ( 18.1 %) | 1192 ( 19.8 %) | 1119 ( 18.9 %) |
| Diabetes Mellitus, n (%) | 32720 ( 12.5 %) | 9939 ( 16.6 %) | 7880 ( 16.4 %) | 998 ( 16.6 %) | 1061 ( 17.9 %) |
| Hypertension, n (%) | 63907 ( 24.5 %) | 20888 ( 34.8 %) | 16638 ( 34.6 %) | 2187 ( 36.4 %) | 2063 ( 34.8 %) |
| Coronary Artery Disease, n (%) | 31495 ( 12.1 %) | 13019 ( 21.7 %) | 10421 ( 21.7 %) | 1293 ( 21.5 %) | 1305 ( 22.0 %) |
| Stroke, n (%) | 8497 ( 3.3 %) | 3392 ( 5.7 %) | 2697 ( 5.6 %) | 352 ( 5.9 %) | 343 ( 5.8 %) |
| Renal Disease, n (%) | 14180 ( 5.4 %) | 5683 ( 9.5 %) | 4475 ( 9.3 %) | 640 ( 10.6 %) | 568 ( 9.6 %) |
| **Procedure Type** | | | | | |
| Cardiovascular, n (%) | 36706 ( 14.0 %) | 24882 ( 41.5 %) | 19840 ( 41.3 %) | 2595 ( 43.2 %) | 2447 ( 41.3 %) |
| Intraperitoneal; intrathoracic; suprainguinal vascular, n (%) | 35741 ( 13.7 %) | 10843 ( 18.1 %) | 8679 ( 18.1 %) | 1074 ( 17.9 %) | 1090 ( 18.4 %) |
| Insulin use prior to admission, n (%) | 12264 ( 4.7 %) | 3762 ( 6.3 %) | 3022 ( 6.3 %) | 366 ( 6.1 %) | 374 ( 6.3 %) |
| Creatinine > 2 mg/dL, n (%) | 10160 ( 3.9 %) | 5602 ( 9.3 %) | 4408 ( 9.2 %) | 637 ( 10.6 %) | 557 ( 9.4 %) |
| RCRI > 2, n (%) | 6285 ( 2.4 %) | 3489 ( 5.8 %) | 2784 ( 5.8 %) | 327 ( 5.4 %) | 378 ( 6.4 %) |
| **Outcomes** | | | | | |
| Death during admission, n (%) | 1452 ( 0.6 %) | 1065 ( 1.8 %) | 865 ( 1.8 %) | 109 ( 1.8 %) | 91 ( 1.5 %) |
| Cardiovascular Mortality, n (%) | 42 ( 0.0 %) | 32 ( 0.1 %) | 21 ( 0.0 %) | 7 ( 0.1 %) | 4 ( 0.1 %) |
| Any major cardiovascular complication, n (%) | 2458 ( 0.9 %) | 1730 ( 2.9 %) | 1400 ( 2.9 %) | 190 ( 3.2 %) | 140 ( 2.4 %) |
| Cardiac Arrest, n (%) | 373 ( 0.1 %) | 227 ( 0.4 %) | 182 ( 0.4 %) | 27 ( 0.4 %) | 18 ( 0.3 %) |
| Myocardial Infarction, n (%) | 391 ( 0.1 %) | 253 ( 0.4 %) | 203 ( 0.4 %) | 30 ( 0.5 %) | 20 ( 0.3 %) |
| Heart Block, n (%) | 204 ( 0.1 %) | 160 ( 0.3 %) | 129 ( 0.3 %) | 22 ( 0.4 %) | 9 ( 0.2 %) |
| Pulmonary Edema, n (%) | 70 ( 0.0 %) | 48 ( 0.1 %) | 40 ( 0.1 %) | 4 ( 0.1 %) | 4 ( 0.1 %) |

**Supplemental Table 8. Most common procedures based on frequency of pre-procedural ECGs by institution**

| CSMC | SHC | CUMC |
|---|---|---|
| CVIC-PCI | Resection of apical lung tumor (eg, Pancoast tumor), including chest wall resection, rib(s) resection(s), neurovascular dissection, when performed; with chest wall reconstruction | CORONARY ARTERY BYPASS GRAFT (CORONARY ARTERY BYPASS GRAFT) |
| CVIC-AORTIC VALVE REPLACEMENT/TRANSFEMORAL | Incision and drainage, open, of deep abscess (subfascial), posterior spine; cervical, thoracic, or cervicothoracic | INSERTION, VENTRICULAR ASSIST DEVICE |
| AORTIC VALVE REPLACEMENT | Percutaneous vertebral augmentation, including cavity creation (fracture reduction and bone biopsy included when performed) using mechanical device, 1 vertebral body, unilateral or bilateral cannulation (eg, kyphoplasty); thoracic | REPLACEMENT, AORTIC VALVE |
| CORONARY ARTERY BYPASS GRAFTING | Discectomy of spine | CHOLECYSTECTOMY, LAPAROSCOPIC |
| MITRAL VALVE REPAIR MINIMALLY INVASIVE ROBOTIC | Total knee replacement | CREATION, TRACHEOSTOMY |
| MITRAL VALVE REPLACEMENT | Fusion or refusion of 2-3 vertebrae | REPAIR OR REPLACEMENT, MITRAL VALVE |
| CVIC-PACEMAKER IMPLANT DUAL | Other exploration and decompression of spinal canal | CREATION, TRACHEOSTOMY, OPEN  AMPUTATION, BELOW KNEE |
| HIP ARTHROPLASTY TOTAL | Total replacement of hip | ARTHROPLASTY, HIP, TOTAL |
| CYSTOSCOPY | Excision or destruction of other lesion or tissue of heart, endovascular approach | INSERTION, SHUNT, VENTRICULOPERITONEAL |
| GI-ESOPHAGOGASTRODUODENOSCOPY WITH PERCUTANEOUS ENDOSCOPIC GASTRO-JEJUNOSTOMY TUBE PLACEMENT | Percutaneous vertebral augmentation, including cavity creation (fracture reduction and bone biopsy included when performed) using mechanical device, 1 vertebral body, unilateral or bilateral cannulation (eg, kyphoplasty); lumbar | STENT DES - CORONARY |